\documentclass[twocolumn,aps,pra,groupedaddress,showpacs]{revtex4}
\usepackage{epsfig,amssymb,amsmath}
\def\comment#1{}\def\labell#1{\label{#1}}
\def\cn{{\cal N}}\def\ce{{\cal E}}\def\ms{{\mathbb S}}\def\mw{{\mathbb W}}
\begin{document}
%{\scriptsize Eprint: quant-ph/0404037}
%Title of paper
\title{Minimum R\'enyi and Wehrl entropies at the output of bosonic
   channels} \author{Vittorio Giovannetti,$^1$\footnote{Now with 
NEST-INFM \& Scuola
Normale Superiore, Piazza dei Cavalieri 7, I-56126, Pisa, Italy.} Seth
Lloyd,$^{1,2}$
   Lorenzo Maccone,$^1$\footnote{Now with QUIT - Quantum Information 
Theory Group,
Dipartimento di Fisica ``A. Volta''
Universita' di Pavia, via A. Bassi 6
I-27100, Pavia, Italy.} Jeffrey H.  Shapiro,$^1$ and Brent J.
   Yen$^1$}\affiliation{$^1$Massachusetts Institute of Technology --
   Research Laboratory of Electronics\\$^2$Massachusetts Institute of
   Technology -- Department of Mechanical Engineering\\ 77
   Massachusetts Ave., Cambridge, MA 02139, USA}
%\date{\today}

\begin{abstract}
  The minimum R\'enyi and Wehrl output entropies are found for bosonic
  channels in which the signal photons are either randomly displaced
  by a Gaussian distribution (classical-noise channel), or in which
  they are coupled to a thermal environment through lossy propagation
  (thermal-noise channel).  It is shown that the R\'enyi output
  entropies of integer orders $z\geqslant 2$ and the output Wehrl
  entropy are minimized when the channel input is a coherent state.
\end{abstract}
 \maketitle

A principal aim of the quantum theory of information is to determine
the ultimate limits on communicating classical
information, i.e., limits arising from quantum 
physics~\cite{chuang,shor}. Among
the various figures of merit employed in this undertaking, one of the 
most basic
is the minimum output entropy~\cite{shorequiv}. It measures the amount
of noise accumulated during the transmission, and may be used to 
derive important
properties, such as the additivity, of other figures of merit, e.g., 
the channel
capacity. Here we will focus  on the R\'enyi and Wehrl output 
entropies for a class
of Gaussian bosonic channels in which the input field undergoes a random
displacement. The R\'enyi entropies
$\{\,S_z(\rho): 0<z<\infty, z\neq 1\,\}$ are a family of functions that
describe the purity of a state~\cite{asomov}. In particular,
the von Neumann entropy $S(\rho)$ can be found from this family, because
$S(\rho) = \lim_{z\rightarrow 1} S_z(\rho)$.  So too can the 
linearized entropy,
because it is a monotonic function of the second-order R\'enyi
entropy~\cite{zyc}. On the other hand, the Wehrl entropy characterizes the
phase-space localization of a bosonic state: its minimum value is realized by
coherent states, whose quadratures have minimum uncertainty product and minimum
uncertainty sum.  In this respect, the Wehrl output entropy can be used to
quantify the channel noise by measuring the phase-space ``spread'' 
of the output
state (see also~\cite{anderson} for a previous analysis of  Wehrl output
entropy). For the classical-noise and thermal-noise channels that we will
consider, we show that coherent-state inputs minimize the R\'enyi 
output entropies
of integer orders $z\geqslant 2$, and the Wehrl output entropy.  The results
presented in this paper are connected with the study of the von Neumann output
entropies of the classical-noise and thermal-noise channels given 
in~\cite{von},
and with the analysis of these channels' additivity properties given
in~\cite{futuro}.

In Sec.~\ref{s:ch} we introduce the classical-noise channel map. In
Sec.~\ref{s:renyi} we analyze the R\'enyi entropy at the output of 
this channel. We
first show that a coherent-state input minimizes $S_z(\rho)$ for 
$z\geqslant 2$ an
integer, and that it minimizes $S_z(\rho)$ for all $z$ when the input is
restricted to be a Gaussian state (Sec.~\ref{s:conjr}).  We then provide lower
bounds, for arbitrary input states, that are consistent with 
coherent-state inputs
minimizing R\'enyi output entropies of all orders (Sec.~\ref{s:other}). In
Sec.~\ref{s:wehrl}, we analyze the Wehrl output entropy, proving that it too is
minimized by  coherent-state inputs.  Moreover, in Sec.\ref{s:rw}, we 
introduce the
R\'enyi-Wehrl entropies, and show that here as well coherent-state inputs yield
minimum-entropy outputs.  The preceding results will all be developed for the
classical-noise channel; in Sec.~\ref{s:th} we show that they also apply to the
thermal-noise channel.

\section{Classical-noise channel}\labell{s:ch}
The classical-noise channel is a unital Gaussian map, i.e., it
transforms Gaussian input states into Gaussian output states while leaving the
identity operator unaffected. It is given by the completely-positive (CP) map
\begin{eqnarray}
{\cal N}_n (\rho)=\int {\rm d}^2 \mu \;
P_{n}(\mu) \; D(\mu)\rho D^{\dagger}(\mu)
\labell{due}
\end{eqnarray}
where
\begin{eqnarray}
P_n(\mu)=\frac{e^{-|\mu|^2/n}}{\pi n},
\labell{tre}
\end{eqnarray}
and $D(\mu)\equiv \exp(\mu a^\dag - \mu^* a)$ is the displacement
operator of the electromagnetic mode $a$ used for the communication.
This map describes a bosonic field that picks up noise through
random displacement by a Gaussian probability distribution $P_n(\mu)$. It is
useful, among other things, to study the fidelity obtainable in
continuous-variable teleportation with finite two-mode
squeezing~\cite{caves1}. Moreover, this simple one-parameter map can
be used to derive properties of more complicated channels, such as the
thermal-noise CP map of Sec.~\ref{s:th}.  When $\cn_n$ acts on a 
vacuum-state input
it produces the thermal-state output
\begin{eqnarray}
\rho'_0\equiv \cn_n(|0\rangle\langle 0|)=\frac{1}{n+1}\left(\frac{n}{n+1}
\right)^{a^\dag a}
\labell{vacuum}\;.
\end{eqnarray}
The covariance property of $\cn_n$ under
displacement implies that a coherent-state input $|\alpha\rangle$
produces the output state
$\rho_\alpha'=D(\alpha)\rho_0'D^\dag(\alpha)$.  See~\cite{hal1l,von,futuro} 
for a more
detailed description of the classical-noise map.
\section{R\'enyi entropies}\labell{s:renyi}
The quantum R\'enyi
entropy $S_{z}(\rho)$ is defined as follows~\cite{zyc},
\begin{eqnarray}
S_{z}(\rho)\equiv-\frac {\ln\mbox{Tr}[\rho^z]}{z-1}\quad
\mbox{for $0<z<\infty,$ $z\neq 1$}
\;\labell{defrenyi},
\end{eqnarray}
It is a monotonic function of the ``${z}$-purity'' Tr$[\rho^{z}]$,
and it reduces to the von Neumann entropy in the limit ${z}\to 1$, viz.,
\begin{equation}
\lim_{z\rightarrow 1}S_z(\rho) = S(\rho) \equiv -\mbox{Tr}[\rho\ln \rho].
\end{equation}
For
$z=2$, the R\'enyi entropy is a monotonic function of the linearized
entropy $S_{\rm lin}(\rho) \equiv 1-$Tr$[\rho^2]$.

We are interested in the
minimum value that $S_z(\rho)$ achieves at the output of the
classical-noise channel, i.e.,
\begin{eqnarray}
\ms_{z}(\cn_n)\equiv\min_{\rho\in{\cal H}}\:S_{z}(\cn_n(\rho))
\;\labell{defmsr},
\end{eqnarray}
where the minimization is performed over all states in the
Hilbert space $\cal H$ associated with the channel's input. The concavity of
$S_{z}$ implies that the minimum in Eq.~(\ref{defmsr}) is achieved by 
a pure-state
input, $\rho=|\psi\rangle\langle\psi|$. Our working hypothesis is that
$S_z(\cn_n(\rho))$ achieves its minimum value when the input is a coherent
state $|\alpha\rangle$, in which case we find that
\begin{eqnarray}
S_{z}(\cn_n(|\alpha\rangle\langle\alpha|))=\frac{\ln[(n+1)^{z}-n^{z}]}{{z}-1}
\;\labell{conjr1}.
\end{eqnarray}
[Note that this quantity does not depend on $\alpha$, thanks to the
invariance of the R\'enyi entropy under unitary transformations.]
Clearly, Eq.~(\ref{conjr1}) provides an upper bound on $\ms_z(\cn_n)$.  We
conjecture that it is also a lower bound, whence
\begin{eqnarray}
\ms_{z}(\cn_n)=\frac{\ln[(n+1)^{z}-n^{z}]}{{z}-1}
\;\labell{conjr}.
\end{eqnarray}
The monotonicity of $S_{z}(\rho)$ with respect to the
${z}$-purity permits restating the conjecture~(\ref{conjr}) as follows,
\begin{eqnarray}
\mbox{Tr}\left\{[\cn_n(\rho)]^{z}\right\}\leqslant
\frac 1{(n+1)^{z}-n^{z}}\;\labell{fine1},
\end{eqnarray}
where the right-hand side of the inequality is the ${z}$-purity at 
the output of
the classical-noise channel when its input is a coherent state. In
Sec.~\ref{s:conjr} we will show that this relation is true for integer
${z}\geqslant 2$, thus proving the conjecture~(\ref{conjr}) in this
case~\cite{newnota}. There we also show that (\ref{fine1}) holds for all
$0<z<\infty$, $z\neq 1$ when the input is restricted to be a Gaussian 
state.  In
Sec.~\ref{s:other} we will present some lower bounds on the R\'enyi 
output entropy
of arbitrary order.

\subsection{Integer-${z}$ R\'enyi entropy}\labell{s:conjr}
 From the definition of the
classical-noise channel, we see that
\begin{eqnarray}
&&\mbox{Tr}\left\{[\cn_n(\rho)]^{k}\right\}=
\int {\rm d}^2\mu_1\cdots {\rm d}^2\mu_{k}\,
P_n(\mu_1)\cdots P_n(\mu_{k})\nonumber\\
&&\times\mbox{Tr}[D(\mu_1)\rho D^\dag(\mu_1)D(\mu_2)\rho
D^\dag(\mu_2)\cdots D^\dag(\mu_{k})]
\;\labell{rhoalla},
\end{eqnarray}
with $k\geqslant 1$ an integer.  For a pure-state input 
$|\psi\rangle$, the trace
can be expressed as
\begin{widetext}
\begin{eqnarray}
\labell{do}
\mbox{Tr}[D(\mu_1)\rho D^\dag(\mu_1)\cdots D^\dag(\mu_{k})]
&=&\langle\psi|D^\dag(\mu_1)D(\mu_2)|\psi\rangle
\langle\psi|D^\dag(\mu_2)D(\mu_3)|\psi\rangle\cdots
\langle\psi|D^\dag(\mu_{k})D(\mu_1)|\psi\rangle\\
&=&\mbox{Tr}\left\{(\rho\otimes\rho\otimes
\cdots\otimes\rho)\left[D_1^\dag(\mu_1)
D_1(\mu_2)\otimes D_2^\dag(\mu_2)
D_2(\mu_3)\otimes
\cdots \otimes D_{k}^\dag(\mu_{k})D_{k}(\mu_1)\right]\right\}
\;,\nonumber
\end{eqnarray}
\end{widetext}
where the ${k}$ scalar products in the input Hilbert space $\cal H$ in
the first line were replaced with a single expectation value on the 
tensor-product
Hilbert space
${\cal H}^{\otimes{k}}$ in the second line. Here
$D_j(\mu)$ is a displacement operator that acts on the $j$th
annihilation operator $a_j$ of this enlarged Hilbert space. With this
replacement, Eq.~(\ref{rhoalla}), which is nonlinear in $\rho$,
can be evaluated as the linear expectation value of an operator $\Theta$ on
${\cal H}^{\otimes{k}}$, i.e.,
\begin{eqnarray}
\mbox{Tr}\left\{[\cn_n(\rho)]^{k}\right\}
=\mbox{Tr}[(\rho\otimes\cdots\otimes\rho)\Theta]
\;\labell{mi},
\end{eqnarray}
with $\Theta$ being a convolution of tensor products of the displacements
$\{D_j\}$, namely
\begin{eqnarray}
\Theta=\int \frac{{\rm d}^2\vec\mu}{(\pi n)^{k}}
\:{e^{-\vec\mu\cdot
C\cdot\vec\mu\,^\dag+\vec\mu\cdot
G^\dag\cdot\vec{a}\,^\dag-\vec{a}\cdot G
\cdot\vec\mu\,^\dag}}
\;\labell{re},
\end{eqnarray}
where $\vec\mu$ is the complex vector $(\mu_1,\cdots,\mu_{k})$ and
$\vec a\equiv(a_1,\cdots,a_{k})$. In Eq.~(\ref{re}),
$C\equiv\frac\openone n+\frac A2$ and $G$ are ${k}\times{k}$
real matrices, with $\openone$ being the identity and
\begin{eqnarray}
A&\equiv&\left[\begin{array}{rrrrrr}
0&-1&0&\cdots&0&1\cr
1&0&-1&\cdots&0&0\cr
0&1&0&\cdots&0&0\cr
\vdots&&&\ddots&&\cr
0&0&0&\cdots&0&-1\cr
-1&0&0&\cdots&1&0\cr
\end{array}
\right]
\;\labell{defa},\\
G&\equiv&\left[\begin{array}{rrrrrr}
-1&1&0&\cdots&0&0\cr
0&-1&1&\cdots&0&0\cr
0&0&-1&\cdots&0&0\cr
\vdots&&&\ddots&&\cr
0&0&0&\cdots&-1&1\cr
1&0&0&\cdots&0&-1\cr
\end{array}
\right]
\;\labell{defg}.
\end{eqnarray}
[The matrix $A$ is null when $k=2$.]  $A$ and $G$ are commuting
circulant matrices~\cite{circulant}, hence they possess a common
basis of orthogonal eigenvectors. This means that there exists a
unitary matrix $Y$ such that $D\equiv Y\:C\:Y^\dag$ and $E\equiv
Y\:G\:Y^\dag$ are diagonal. Rewriting $\Theta$ from
Eq.~(\ref{re}) in factored form by performing the change of
integration variables $\vec\nu\equiv\vec\mu\cdot Y^\dag$, and then
introducing the new annihilation operators $\vec b\equiv \vec a\cdot
Y^\dag$, we find that
\begin{eqnarray}
\Theta=\bigotimes_{j=1}^{k} \Theta_j
\;\labell{fatt},
\end{eqnarray}
with
\begin{eqnarray}
\Theta_j\equiv\frac 1{n|e_j|^2}\int \frac{{\rm d}^2\nu}{\pi}
\:{e^{-d_j|\nu|^2/|e_j|^2}}\:D_{b_j}(\nu)
\;\labell{re2},
\end{eqnarray}
where $D_{b_j}(\nu)\equiv\exp[\nu b_j^\dag-\nu^*b_j]$ is the
displacement operator associated with $b_j$, while $d_j$ and $e_j$ are
the $j$th diagonal elements of the matrices $D$ and $E$, respectively
(i.e., they are the $j$th eigenvalues of $C$ and $G$). As discussed in
App.~\ref{s:ren}, the operator $\Theta_j$ is diagonal in the Fock
basis of the mode $b_j$ and takes the thermal-like form~\cite{caves1}
\begin{eqnarray}
\Theta_j=\frac{2/n}{2d_j+|e_j|^2}\left(\frac{2d_j-|e_j|^2}{2d_j+|e_j|^2}\right)^{b_j^\dag
   b_j}
\;\labell{defqj}.
\end{eqnarray}
Because the $\{d_j\}$ have positive real parts equal to $1/n$ [see
Eq.~(\ref{reale})], the vacuum state of $b_j$ is the $\Theta_j$-eigenvector
whose associated eigenvalue has the maximum absolute value,
${2/[n(2d_j+|e_j|^2)]}$.  It then follows from Eq.~(\ref{fatt})
that for any state $R\in{\cal H}^{\otimes{k}}$ we have
\begin{eqnarray}
\Big|\mbox{Tr}[R\:\Theta]\Big|&\leqslant&\prod_{j=1}^{k}
\frac{2/n}{2d_j+|e_j|^2}=\frac{1/n^{k}}
{\det[C+G^\dag G/2]} \nonumber\\&=&
\frac 1{(n+1)^{k}-n^{k}}
\;\labell{croma},
\end{eqnarray}
where in deriving the first equality we have used the invariance of
the determinant under the unitary transformation $Y$. Because
inequality~(\ref{fine1}) now follows directly from Eq.~(\ref{mi}),
this completes the proof: for integer $k\geqslant 2$ the maximum
$k$-purity (or, equivalently, the minimum R\'enyi entropy
$\ms_k(\cn_n)$) is provided by a coherent state input (see also
App.~\ref{s:emis}).
\paragraph*{Gaussian-state inputs:  }
Suppose that the channel input is restricted to be a Gaussian state, 
$\rho_G$.  It
is easy to show that a coherent-state input minimizes $S_z(\rho_G)$ for all
$0<z<\infty$, $z\neq 1$.  A Gaussian state is completely characterized by its
mean $\langle a\rangle$ and its covariance matrix,
\begin{eqnarray}
\Gamma\equiv\left[\begin{array}{cc}
\langle\{\Delta a,\Delta a^\dag\}\rangle/2&
\langle(\Delta a)^2\rangle\cr
\langle(\Delta a)^2\rangle&
\langle\{\Delta a,\Delta a^\dag\}\rangle/2
\end{array}\right]
\;\labell{correlation},
\end{eqnarray}
where $\langle\;\cdot\;\rangle\equiv \mbox{ Tr}[\;\cdot\;\rho_G]$ is
expectation with respect to $\rho_G$, $\Delta a \equiv a-\langle
a\rangle$, and
$\{\;\cdot\;,\;\cdot\;\}$ denotes the anticommutator.  As shown in \cite{von},
the classical-noise channel's output state $\rho'_G$, when its input 
is $\rho_G$,
is also Gaussian.   The mean, $\langle a\rangle$, is unaffected by the CP map
${\cal N}_n$, but the covariance matrix is modified by the presence of
classical noise, viz., $\Gamma \rightarrow \Gamma' = \Gamma + n\openone$.

By concatenating two unitary
transformations---a displacement to drive
$\langle a\rangle$ to zero, and a squeeze operator to symmetrize the quadrature
uncertainties---$\rho'_G$ can be converted into the thermal state
\begin{eqnarray}
\tau'_G =\frac{1}{n'+1}\left(\frac{n'}{n'+1}
\right)^{a^\dag a}
\labell{Gthermal}\;,
\end{eqnarray}
where $n' = \sqrt{\det\Gamma'} -1/2$.  The state (\ref{Gthermal}) has R\'{en}yi
entropy
\begin{equation}
S_z(\tau'_G) = \frac{\ln[(n'+1)^{z}-{n'}^{z}]}{{z}-1} \quad
\mbox{for $0<z<\infty$, $z\neq 1$}.
\labell{RenyiG}
\end{equation}
Moreover, because R\'{e}nyi entropy is
invariant under unitary transformations, we have $S_z(\rho'_G) = S_z(\tau'_G)$.
Equation~(\ref{RenyiG}) thus shows that $S_z(\rho'_G)$ is 
monotonically increasing
with increasing $n' = \sqrt{\det\Gamma'}-1/2$, and in \cite{von} we showed that
$\min_{\rho_G}( \sqrt{\det \Gamma'}-1/2) = n$ is achieved by coherent-state
inputs.  It follows that $S_z({\cal N}_n(\rho_G))$ is minimized, for all $0 <z
<\infty$,
$z\neq 1$, when the channel input is a coherent state.  The corresponding
Gaussian-state result for the von Neumann entropy at the 
classical-noise channel's
output was derived in~\cite{von}.

\paragraph*{Comments:  }
The most interesting cases for integer-order R\'{e}nyi output entropy are
$k=2$ and
$k\to\infty$, where we have
\begin{eqnarray}
\ms_2&=&\ln(2n+1)\;\labell{sdue},\\
\ms_\infty&=&\ln(n+1)\;\labell{infty}.
\end{eqnarray}
Equation~(\ref{sdue}) has been used in~\cite{von} to derive lower
bounds for the von Neumann entropy at the output of the 
classical-noise channel. On
the other hand, Eq.~(\ref{infty}) establishes an upper bound on the
maximum eigenvalue $\lambda_{\rm max}$ of any output state $\cn_n(\rho)$
of the channel. This is so because the R\'enyi
entropy becomes $S_\infty(\cn_n(\rho))=-\ln(\lambda_{\rm max})$ in the limit
${k}\to\infty$~\cite{zyc}, and Eq.~(\ref{infty}) requires that
$\lambda_{\rm max}\leqslant1/(n+1)$.

\subsection{R\'enyi entropy lower bounds}\labell{s:other}
In this section we develop four lower bounds on $\ms_z$ for arbitrary 
$z$, which
support the conjecture~(\ref{conjr}).
%da renyi.lb.f
\begin{figure}[hbt]
\begin{center}
\epsfxsize=.8
\hsize\leavevmode\epsffile{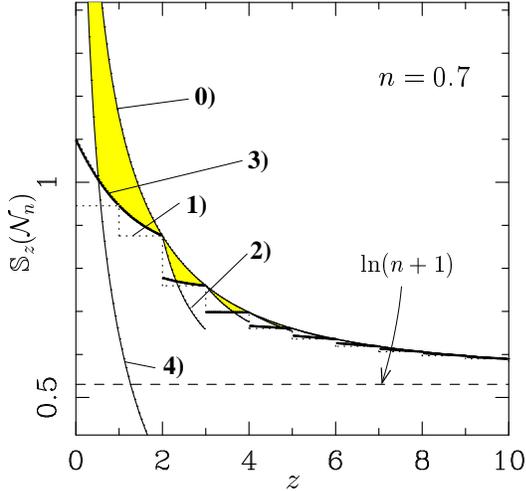}
\end{center}
%\vspace{-.5cm}
\caption{Bounds on the minimum R\'enyi entropies $\ms_{z}(\cn_n)$ (in nats)
  as a function of ${z}$: $\ms_{z}$ is restricted to the gray region.
  The upper bound {\bf 0)} is the the R\'enyi output entropy of the
  vacuum input, i.e., the right-hand side of Eq.~(\ref{conjr}).  Lower
  bound~{\bf 1)} (dotted staircase) derives from the fact that $S_{z}$
  is a decreasing function of ${z}$. For ${z}>1$, it follows from
  (\ref{lb1}) and the integer-${z}$ minimum R\'enyi entropy calculated
  in Sec.~\ref{s:conjr}; for ${z}\leqslant 1$, it is equal to
  $\bar\ms(\cn_n)$, the best of the the von Neumann output entropy
  lower bounds from~\cite{von} (in this case $\bar\ms\sim0.56$).
  Lower bounds {\bf 2)} and {\bf 4)} are given by (\ref{lb2})
  and~(\ref{lb4}), respectively.  Lower bound~{\bf 3)} (thick line) is
  the greater of (\ref{lb3}) and (\ref{lb3p}).  For high values of
  $z$, the maximum of lower bounds {\bf 1)}--{\bf 4)} asymptotically
  coincides with the upper bound {\bf 0)}; in the figure lower bound
  {\bf 2)} becomes indistinguishable from the upper bound once
  ${z}\gtrsim 5$.  The dashed line is $S_{\infty} = \ln(n+1)$.  }
\labell{f:ren}\end{figure}

%da renyi.lb.f
\begin{figure}[hbt]
\begin{center}
\epsfxsize=.8
\hsize\leavevmode\epsffile{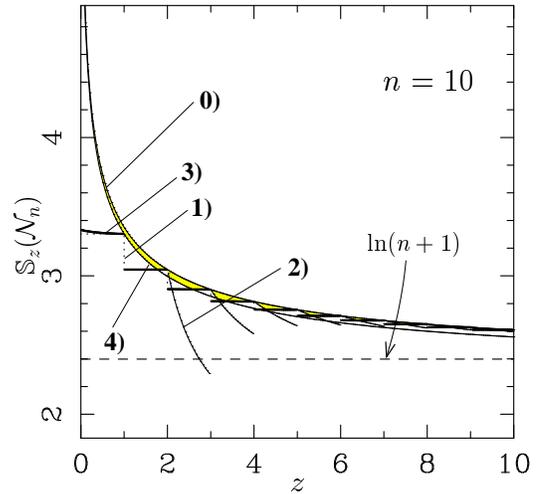}
\end{center}
%\vspace{-.5cm}
\caption{Same as Fig.~\ref{f:ren}, but for a different value of the noise
   parameter $n$. Note that the lower bounds approach the upper bound
   {\bf 0)} for high values of $n$, indicating that our conjecture is 
asymptotically
   true.  } \labell{f:ren1}\end{figure}

\paragraph*{Lower bound {\bf 1)}:}
The R\'enyi entropy $S_{z}(\rho)$ is a decreasing function of ${z}$~\cite{zyc}.
So, using our knowledge of $\ms_k(\cn_n)$ for integers $k\geqslant 
2$, we have that
\begin{eqnarray}
\ms_{z}(\cn_n)\geqslant\ms_k(\cn_n)=\frac{\ln((n+1)^k-n^k)}{k-1}
\;\labell{lb1},
\end{eqnarray}
for all ${z}\leqslant k$.
For ${z}\leqslant 1$, we can employ the best of the von Neumann output entropy
lower bounds that we established in~\cite{von} to
derive a tighter lower bound on the R\'enyi entropy.  Together with
Eq.~(\ref{lb1}), this additional bound produces the staircase function {\bf 1)}
shown in Figs.~\ref{f:ren} and~\ref{f:ren1}.
\paragraph*{Lower bound {\bf 2)}:}
The definition of the R\'enyi entropy leads to the following
monotonicity property~\cite{zyc},
\begin{eqnarray}
\frac{{z}-1}{{z}}S_{{z}}(\rho)\geqslant
\frac{{z}'-1}{{z}'}S_{{z}'}(\rho)
\;\labell{pro},
\end{eqnarray}
for any ${z}\geqslant{z}'$ and for all $\rho$. Allowing $\rho$
to be an arbitrary output state from the channel $\cn_n$, and minimizing both
sides of (\ref{pro}) over all the possible inputs, we obtain
\begin{eqnarray}
\frac{{z}-1}{{z}}\ms_{{z}}(\cn_n)\geqslant
\frac{{z}'-1}{{z}'}\ms_{{z}'}(\cn_n)
\;\labell{pro1}.
\end{eqnarray}
When ${z}'=k\geqslant 2$ is an integer, this relation provides the lower
bound
\begin{eqnarray}
\ms_{z}(\cn_n)\geqslant\frac{z}{{z}-1}\frac{\ln((n+1)^k-n^k)}k
\;\labell{lb2},
\end{eqnarray}
for $z\geqslant k$, which is shown as curve {\bf 2)} in Figs.~\ref{f:ren}
and~\ref{f:ren1}.
\paragraph*{Lower bound {\bf 3)}:}
Using the relation between different measures of entropy established
in~\cite{renyi,wang}, the following inequality can be derived
(see App.~\ref{s:lb3}):
\begin{eqnarray}
\ms_z(\cn_n)\geqslant-\frac 1{z-1}\ln\left\{h_z\left[h_{k}^{-1}\left(
\frac 1{(n+1)^k-n^k}\right)\right]\right\}
\;\labell{lb3},
\end{eqnarray}
for all $z\leqslant k$ and integers $k\geqslant 2$. Here, $h_z(x)$ is
the function defined in~(\ref{hdef}) and $h^{-1}_z(x)$ its inverse.
For $z\leqslant 1$ a further lower bound can be obtained from
$\bar\ms(\cn_n)$, the best of the lower bounds on the von Neumann 
output entropy
given in~\cite{von}:
\begin{eqnarray}
\ms_z(\cn_n)\geqslant-\frac 1{z-1}\ln\left\{h_z\left[v^{-1}\left(
\bar\ms(\cn_n)\right)\right]\right\}
\;\labell{lb3p},
\end{eqnarray}
where $v^{-1}(x)$ is the inverse of the function $v(x)$ defined in
Eq.~(\ref{vdef}).  Curve {\bf 3)} of Figs.~\ref{f:ren}
and~\ref{f:ren1} has been obtained by considering the maximum of all
the functions on the right-hand sides of (\ref{lb3})
and~(\ref{lb3p}).

\paragraph*{Lower bound {\bf 4)}:}
Our final lower bound can be derived from the inequality~\cite{von}
\begin{eqnarray}
\mbox{Tr}\{[\cn_n(\rho)]^{z}\}\leqslant\frac
{\mbox{Tr}[\cn_{n/{z}}(\rho)]}
{{z}\:n^{{z}-1}}=\frac 1{{z}\: n^{{z}-1}}
\;\labell{boh1},
\end{eqnarray}
for ${z}\geqslant 1$,
which implies
\begin{eqnarray}
S_{z}(\cn_n(\rho))\geqslant\frac{\ln{z}}{{z}-1}+\ln n
\;\labell{lb4},
\end{eqnarray}
for any input $\rho$ and ${z}\geqslant 1$. Inequality~(\ref{boh1}) was derived
in~\cite{von} from the {\it convexity} of $x^z$ for $z\geqslant 1$.  For
${z}\leqslant 1$, the function $x^{z}$ is {\it concave} and we obtain
\begin{eqnarray}
\mbox{Tr}\{[\cn_n(\rho)]^{z}\}\geqslant\frac
{\mbox{Tr}[\cn_{n/{z}}(\rho)]}
{{z}\:n^{{z}-1}}=\frac 1{{z}\: n^{{z}-1}}
\;\labell{bohR}.
\end{eqnarray}
The sign change associated with the
$1/({z}-1)$ factor in the R\'enyi entropy definition then shows that 
(\ref{lb4})
also applies for ${z}\leqslant 1$.  Lower bound~(\ref{lb4}) is plotted as curve
{\bf 4)} in Figs.~\ref{f:ren} and~\ref{f:ren1}.

\section{Wehrl entropy}\labell{s:wehrl}
The Wehrl entropy is the continuous Boltzmann-Gibbs entropy of the
Husimi probability function for the state $\rho$~\cite{wehrl},
\begin{eqnarray}
{W}(\rho)\equiv-\int {{\rm d}^2\mu}
\:Q(\mu)
\ln [\pi Q(\mu)]
\;\labell{defwehrl},
\end{eqnarray}
where $Q(\mu)\equiv\langle\mu|\rho|\mu\rangle/\pi$ with $|\mu\rangle$
a coherent state. The Wehrl entropy provides a measurement of the
``localization'' of the state $\rho$ in the phase space: its minimum
value is achieved on coherent states~\cite{wehrl,lieb}. It is also
useful in characterizing the statistics associated with heterodyne
detection~\cite{heter}. Here we study this minimum restricted to the
output states from the classical-noise channel, i.e.,
\begin{eqnarray}
\mw(\cn_n)\equiv\min_{\rho\in{\cal H}}\:W(\cn_n(\rho))
\;\labell{defmw}.
\end{eqnarray}
We will show that coherent-state inputs achieve this minimum, which is then
given by
\begin{eqnarray}
\mw(\cn_n)=1+\ln(n+1)
\;\labell{q}.
\end{eqnarray}

The output-state Husimi function $Q'(\mu)$ for the channel map
$\cn_n$ is the convolution of the input-state Husimi function
$Q(\mu)$ with the Gaussian probability distribution $P_n$ from Eq.~(\ref{tre}),
\begin{eqnarray}
Q'(\mu)=(P_n*Q)(\mu)=\int {\rm d}^2\nu\: P_n(\nu)\: Q(\mu-\nu)
\;\labell{husimievol}.
\end{eqnarray}
This property can be used to show that the right-hand side of
Eq.~(\ref{q}) is an upper bound for $\mw$, because it is the value
achieved by a coherent-state input. In particular, the Husimi function of the
coherent state
$|\alpha\rangle$ is
$Q_\alpha(\mu)\equiv\left|\langle\alpha|\mu\rangle\right|^2/\pi=
\exp(-|\mu-\alpha|^2)/\pi$, which evolves into
\begin{eqnarray}
Q'_{\alpha}(\mu)=\frac{\exp\left[-\frac{|\mu-\alpha|^2}{n+1}\right]}{\pi(n+1)}
\;\labell{q11},
\end{eqnarray}
under~(\ref{husimievol}).
The resulting Wehrl output entropy is then
\begin{eqnarray}
W(\cn_n(|\alpha\rangle\langle\alpha|))&=&
\int {\rm d}^2\mu\:Q'_{\alpha}(\mu)\:\frac{|\mu-\alpha|^2}{n+1}+\ln
(n+1)\nonumber\\&=&1+\ln(n+1)
\;\labell{www}.
\end{eqnarray}
(An analogous result was also given in~\cite{hall}.)
To show that this quantity is also a lower bound for $\mathbb W$, we
use Theorem~6 of~\cite{lieb},  which states that for two
probability distributions $f(\mu)$ and $h(\mu)$ on ${\mathbb
   C}$ we have
\begin{eqnarray}
W((f*h)(\mu))&\geqslant&\lambda\;W(f(\mu))+(1-\lambda)W(h(\mu))
\nonumber\\&&-\lambda\ln\lambda-
(1-\lambda)\ln(1-\lambda)
\;\labell{theorem}
\end{eqnarray}
for all $\lambda\in[0,1]$,
where $f*h$ is the convolution of $f$ and $h$ and where the Wehrl
entropy of a probability distribution is found from Eq.~(\ref{defwehrl})
by replacing $Q(\mu)$ with the given distribution.  Choosing $f={P_n}$ and
$h={Q}$ makes $f*h$  the classical-noise channel's output-state 
Husimi function,
$Q'$. Hence,
inequality~(\ref{theorem}) implies that
\begin{eqnarray}
W(\cn_n(\rho))&\geqslant& \lambda\:W(P_n)+(1-\lambda)\:W(\rho)
\nonumber\\&&-\lambda\ln\lambda-
(1-\lambda)\ln(1-\lambda)
\;\labell{a},
\end{eqnarray}
where $W(P_n)=1+\ln n$ is the Wehrl entropy of the distribution $P_n$.
Because $W(\rho)\geqslant 1$ for any $\rho$~\cite{lieb}, Eq.~(\ref{a})
gives
\begin{eqnarray}
W(\cn_n(\rho))\geqslant \lambda\ln n+1
-\lambda\ln\lambda-
(1-\lambda)\ln(1-\lambda)
\labell{d},
\end{eqnarray}
which for $\lambda=n/(n+1)$ becomes
\begin{eqnarray}
W(\cn_n(\rho))&\geqslant& 1+\ln(n+1)
\;\labell{b}.
\end{eqnarray}
Inasmuch as this relation applies for all $\rho$, Eq.~(\ref{q}) then follows.

\subsection{R\'enyi-Wehrl entropies}\labell{s:rw}
The ${z}$--R\'enyi-Wehrl entropies are defined by~\cite{rwen}
\begin{eqnarray}
W_{z}(\rho)&\equiv&-\frac 1{{z}-1}\ln(m_{z}(\rho))\;,
\;\labell{defrw}\\
m_{z}(\rho)&\equiv&\int \frac{{\rm d}^2\mu}{\pi}[\pi Q(\mu)]^{z}\;,
\end{eqnarray}
where $Q(\mu)$ is the Husimi function of $\rho$ and ${z}\geqslant 1$.
Thus, the Wehrl entropy $W(\rho)$ is the limit as ${z}\to 1$ of 
$W_{z}(\rho)$, and
$W_{z}(\rho)$ achieves its minimum value, $\ln(z)/({z}-1)$, when $\rho$ is a
coherent state $|\alpha\rangle$, for which
$m_{z}(|\alpha\rangle\langle\alpha|)=1/{z}$. For arbitrary
$\rho$, Theorem~3 of~\cite{lieb} implies
\begin{eqnarray}
m_{{z}}(\rho)
=\int\frac{{\rm d}^2\mu}{\pi}[\pi\:Q(\mu)]^{{z}}\leqslant
\frac 1{z}
\;.
\end{eqnarray}

We now show that $\mw_z(\cn_n) \equiv \min_\rho(W_z({\cal 
N}_n(\rho))$ is achieved
by coherent-state inputs.  From Eq.~(\ref{q11}), the classical-noise channel's
R\'enyi-Wehrl output entropy for the coherent-state input 
$|\alpha\rangle$ can be
shown to be
\begin{eqnarray}
W_{z}(\cn_n(|\alpha\rangle\langle\alpha|))=
\frac{\ln{z}}{{z}-1}+\ln (n+1)
\;\labell{ok}.
\end{eqnarray}
To show that the right-hand side of this equation is the global
minimum, we observe that, for an arbitrary state $\rho$ and for all
$p,q\geqslant 1$ such that $1/p+1/q=1+1/{z}$, the sharp form of
Young's inequality (Lemma~5 of Ref.~\cite{lieb}) together
Eq.~(\ref{husimievol}) give
\begin{eqnarray}
&&m_{z}(\cn_n(\rho))=
\int \frac{{\rm d}^2\mu}{\pi}[\pi\:Q'(\mu)]^{z}\leqslant
\left(\frac{C_pC_q}{C_{z}}\right)^{2{z}}
\nonumber\\&&\times
\left[\int\frac{{\rm d}^2\mu}{\pi}[\pi\:Q(\mu)]^{p}\right]^{{z}/p}
\left[\int\frac{{\rm d}^2\mu}{\pi}\frac{e^{-q|\mu|^2/n}}{n^q}
\right]^{{z}/q}
\nonumber\\&&
=\left(\frac{C_pC_q}{C_{z}}\right)^{2{z}}
\left[m_p(\rho)\right]^{{z}/p}
\left[\frac n{qn^q}\right]^{{z}/q}
\;\labell{kk},
\end{eqnarray}
where $C_p$, $C_q$, and $C_{z}$ are the Young's inequality constants,
\begin{eqnarray}
C_x\equiv\left[\frac{x^{1/x}}{(x')^{1/x'}}\right]^{1/2}
\;\qquad x'\equiv
x/(x-1)
\;\labell{x}.
\end{eqnarray}
Choosing $p=(n+1){z}/(n{z}+1)$ and, hence, $q=(n+1){z}/({z}+n)$, we
then obtain
\begin{eqnarray}
&&m_{z}(\cn_n(\rho))\leqslant\frac 1{{z}(n+1)^{{z}-1}}
\;,
\end{eqnarray}
which, via Eq.~(\ref{defrw}), completes the proof.

\section{Thermal-noise channel}\labell{s:th}
Thus far we have limited our attention to the CP map $\cn_n$ 
associated with the
classical-noise channel. This channel is a limiting case of the
thermal-noise channel, in which the
signal mode
$a$ and a thermal-reservoir mode $b$ couple to the
channel output through a beam splitter~\cite{von,futuro}.  The thermal-noise
channel's CP map $\ce_\eta^N$ is obtained by tracing away the noise 
mode---which
initially is in a thermal state with average photon number $N$---from the
evolution
\begin{eqnarray}
a\longrightarrow \sqrt{\eta}\; a + \sqrt{1-\eta} \; b
\;\labell{evol},
\end{eqnarray}
where $\eta$ is the coupling parameter (the channel's quantum
efficiency).
A detailed characterization of the two maps $\cn_n$ and
$\ce_\eta^N$ is given in~\cite{von}, where, in particular, it is shown
that they are related through the composition rule
\begin{eqnarray}
{\cal E}_\eta^N(\rho)=\left({\cal N}_{(1-\eta)N}\circ{\cal 
E}_\eta^0\right)(\rho)
\equiv {\cal N}_{(1-\eta)N}\left({\cal E}_\eta^0(\rho)\right)
\;\labell{pr3}.
\end{eqnarray}
This means that the thermal-noise channel $\ce_\eta^N$ can be regarded
as the application of the map $\cn_n$ to the output of the pure-loss
channel ${\cal E}_\eta^0$, with the latter being a zero-temperature
($N=0$) thermal-noise channel.

We can use (\ref{pr3}) to extend all the analyses from the previous
sections to the thermal-noise channel. Specifically, the minimum
$z$-R\'enyi output entropy of the thermal-noise channel, obeys
\begin{eqnarray}
\ms_{z}(\ce_\eta^N)=\ms_{z}(\cn_{(1-\eta)N}\circ\ce_\eta^0)
\geqslant\ms_{z}(\cn_{(1-\eta)N})
\;\labell{sa},
\end{eqnarray}
because the implicit minimization on the left is performed over a
subset of the states considered in the implicit minimization on the
right.  Replacing $n$ with $(1-\eta)N$ in this inequality, we
immediately find that the lower bounds from Sec.~\ref{s:other} also
apply to the thermal-noise channel $\ce_\eta^N$.  Moreover, for
$z\geqslant 2$ an integer, (\ref{sa}) becomes an equality, because the
implicit minimum on the left is achieved by the vacuum-state input
$|0\rangle$, for which, according to Eq.~(\ref{pr3}),
\begin{eqnarray}
\ce_\eta^N(|0\rangle\langle 0|)=\cn_{(1-\eta)N}(|0\rangle\langle 0|)
\;\labell{v}.
\end{eqnarray}
This proves that for integers $k\geqslant 2$ the minimum R\'{e}nyi 
entropy at the
output of the thermal-noise channel is
\begin{eqnarray}
\ms_k(\ce_\eta^N)=\frac{\ln\{[(1-\eta)N+1]^k-[(1-\eta)N]^k\}}{k-1}\;
\labell{w1}.
\end{eqnarray}
Some preliminary results in this regard were obtained in~\cite{paz},
where it was shown that the linearized entropy of the thermal-noise
channel---i.e., $S_2({\cal E}_\eta^N(\rho))$---is minimized by the 
vacuum input in
the limit of low coupling ($\eta
\ll 1)$ and high temperature ($N\gg 1$).

When the input to the thermal-noise channel is a Gaussian state
$\rho_G$ with covariance matrix $\Gamma$, the output state will be
Gaussian with covariance matrix $\Gamma' = \eta\Gamma +
(1-\eta)(N+1/2)\openone$ \cite{von}.  We have previously shown that
$\min_{\rho_G}(\sqrt{\det \Gamma'}-1/2) = (1-\eta)N$ is achieved when
the input is a coherent state, which shows that coherent-state inputs
minimize $S_z({\cal E}_\eta^N(\rho_G))$ for all $0<z<\infty$, $z\neq
1$.

Finally, arguments identical to the ones given earlier for the 
minimum Wehrl and
R\'{e}nyi-Wehrl entropies at the output of the classical-noise channel also
apply to the minimum Wehrl and R\'{e}nyi-Wehrl entropies at the output of
the thermal-noise channel.  Because the minimum values
$\mw(\cn_n)$ and
$\mw_z(\cn_n)$ are achieved by coherent-state inputs, such as the vacuum,
Eqs.~(\ref{pr3}) and~(\ref{v}) imply that
\begin{eqnarray}
\mw(\ce_\eta^N)&=&1+\ln[(1-\eta)N+1]\nonumber\\
\mw_z(\ce_\eta^N)&=&\frac{\ln z}{z-1}+\ln[(1-\eta)N+1]
\;\labell{ww}.
\end{eqnarray}
with these minima being realized by coherent-state inputs.

\section{Conclusion}\labell{s:concl}
The minimum R\'enyi and Wehrl output entropies have been analyzed for bosonic
channels in which the signal photons are disturbed by classical 
additive Gaussian
noise, or by a combination of propagation loss and Gaussian noise. 
We conjectured
that the R\'enyi output entropy is minimized by coherent-state inputs. Some
arguments were provided to place this conjecture on solid ground.  In 
particular,
we have shown that it is true for integer orders greater than
one, and it is true when the input state is restricted to being 
Gaussian.  For the
general case---non-integer orders and arbitrary input states---we have provided
entropic lower bounds that are compatible with the upper bound implied by the
conjecture. In addition, we have shown that coherent-state inputs minimize
the Wehrl and the R\'enyi-Wehrl output entropies for these two channels.

\appendix
\section{Derivation of Eq.~(\ref{defqj})}\labell{s:ren}
In this appendix we show that the operator $\Theta_j$ defined in
Eq.~(\ref{re2}) coincides with the right-hand side of
Eq.~(\ref{defqj}).  The easiest way to prove this assertion is to show that
these operators have the same characteristic function.  We take
advantage of the interesting analysis in~\cite{caves1}, where the
maximal-entanglement teleportation fidelity is calculated for the
classical-noise channel, and $k=2$ version of (\ref{defqj}) was implicitly
demonstrated.

 From Eq.~(\ref{re2}), we immediately see that the symmetrical
characteristic function~\cite{walls} of the operator $\Theta_j$ is
\begin{eqnarray}
\chi_{j}(\nu)\equiv\mbox{Tr}[\Theta_j\:D_j(\nu)]=
\frac{\exp(-{d_j|\nu|^2}/{|e_j|^2})}{n|e_j|^2}
\;\labell{chrqj}.
\end{eqnarray}
On the other hand, the characteristic function of the right-hand side
of Eq.~(\ref{defqj}) is given by
\begin{eqnarray}
&&\chi'_j(\nu)=\frac {2/n}{2d_j+|e_j|^2}\sum_{m=0}^\infty
\left(\frac{2d_j-|e_j|^2}{2d_j+|e_j|^2}\right)^m \langle
m|D_{b_j}|m\rangle\nonumber\\\nonumber
&&=
\frac {2/n}{2d_j+|e_j|^2}\sum_{m=0}^\infty
\left(\frac{2d_j-|e_j|^2}{2d_j+|e_j|^2}\right)^m
e^{-|\nu|^2/2}L_m(|\nu|^2)
\;,\\&&\labell{llag}
\end{eqnarray}
where $\{|m\rangle\}$ are the Fock states of the $b_j$ mode and $L_m$ is the
Laguerre polynomial of order $m$. From the definition of the matrix
$C$ [see Eq.~(\ref{re})] we know that
\begin{eqnarray}
d_j=1/n+i\xi_j\;\labell{reale},
\end{eqnarray}
  where $\{i\xi_j\}$
are the imaginary eigenvalues of the real anti-symmetric matrix $A$ from
Eq.~(\ref{defa}).  This implies that the $\{d_j\}$ have positive real parts, so
that the absolute value of the parenthetical term in Eq.~(\ref{llag}),
$({2d_j-|e_j|^2})/({2d_j+|e_j|^2})$, is less than
one. The summation in Eq.~(\ref{llag}) can thus be performed using the
formula~\cite{grad}
\begin{eqnarray}
\sum_{m=0}^\infty z^mL_m(x)=\frac {\exp[xz/(z-1)]}{1-z}
\qquad\mbox{for }|z|<1\;\labell{lag}.
\end{eqnarray}
With this relation Eq.~(\ref{llag}) yields $\chi_j$, concluding the
derivation.
\subsection*{Examples}
Here, for the sake of clarity, we carry out calculations of the
$\{\Theta_j\}$ for the cases $k=2$ and $k=3$.

When $k=2$, the matrix $A$ is null and $G$ has eigenvalues
$e_1=0$ and $e_2=2$. The unitary transformation that diagonalizes
$A$ and $G$ is then
\begin{equation}
Y=\frac
1{\sqrt{2}}\left[\begin{array}{rr}1&1\cr-1&1\end{array}\right]\;,
\end{equation}
so
that $\Theta_1=\openone$ on the mode $b_1=(a_1+a_2)/\sqrt{2}$, and
\begin{eqnarray}
\Theta_2=\frac 1{1+2n}\left(\frac{1-2n}{1+2n}\right)^{b_2^\dag b_2}
\;\labell{sq},
\end{eqnarray}
on the mode $b_2=(a_2-a_1)/\sqrt{2}$.

When $k=3$, the matrix $A$ has eigenvalues $i\xi_1=0$,
$i\xi_2=i\sqrt{3}/2$, and $i\xi_3=-i\sqrt{3}/2$. On the other hand,
$G$ has eigenvalues $e_1=0$, $e_2=i\sqrt{3}e^{i2\pi/3}$, and
$e_3=-i\sqrt{3}e^{i2\pi/3}$. Now the unitary matrix $Y$ is
\begin{eqnarray}
Y=\frac 1{\sqrt{3}}\left[\begin{array}{ccc}1&1&1\cr
     e^{2i\pi/3}&e^{4i\pi/3}&1\cr
e^{4i\pi/3}&e^{2i\pi/3}&1\end{array}\right]
\;\labell{maty2},
\end{eqnarray}
so that $\Theta_1=\openone$ on the mode $b_1=(a_1+a_2+a_3)/\sqrt{3}$,
\begin{eqnarray}
\Theta_2=\frac
2{2+(3+i\sqrt{3})n}\left(\frac{{2+(-3+i\sqrt{3})n}}{{2+(3+i\sqrt{3})n}}\right)^{b_2^\dag
   b_2}
\;\labell{sq1},
\end{eqnarray}
on the mode $b_2=(e^{4i\pi/3}a_1+e^{2i\pi/3}a_2+a_3)/\sqrt{3}$, and
\begin{eqnarray}
\Theta_3=\frac
2{2+(3-i\sqrt{3})n}\left(\frac{{2-(3+i\sqrt{3})n}}{{2+(3-i\sqrt{3})n}}\right)^{b_3^\dag
   b_3}
\;\labell{sq12},
\end{eqnarray}
on the mode $b_3=(e^{2i\pi/3}a_1+e^{4i\pi/3}a_2+a_3)/\sqrt{3}$.

\section{Entropy-minimizing input states  }\labell{s:emis}
Even though it was already proven in Sec.~\ref{s:renyi} [see
Eqs.~(\ref{conjr1}) and~(\ref{fine1})], it is instructive to use a
different method to explicitly show that the upper bound~(\ref{croma})
on the integer-order R\'enyi output entropy can be achieved by
employing a vacuum-state input, $\rho=|0\rangle\langle 0|$. By
construction, the vacuum state for the $b_j$ modes,
$R_0=|0\rangle_{b_1}\langle
0|\otimes\cdots\otimes|0\rangle_{b_{k}}\langle 0|$, saturates this
bound. Because $\vec a$ is obtained from $\vec b$ through the unitary
matrix $Y$, the state $R_0$ is also the vacuum state of the $\vec a$
modes. Indeed, from the symmetric characteristic function
decomposition, we find
\begin{eqnarray}
R_0&=&\int \frac{{\rm d}^2\vec\nu}{\pi^{k}}\:\exp[-|\vec\nu|^2/2
+\vec\nu\cdot\vec b\,^\dag-\vec b\cdot\vec\nu\,^\dag]\nonumber\\
&=&\int \frac{{\rm d}^2\vec\mu}{\pi^{k}}\:\exp[-|\vec\mu|^2/2
+\vec\mu\cdot\vec a\,^\dag-\vec a\cdot\vec\mu\,^\dag]\nonumber\\
&=&|0\rangle_{a_1}\langle
0|\otimes\cdots\otimes|0\rangle_{a_{k}}\langle 0|
\;\labell{vuoto},
\end{eqnarray}
where $\vec\nu=\vec\mu\cdot Y^\dag$.  From Eq.~(\ref{conjr1}) we know
that all coherent-state inputs produce the same R\'enyi output 
entropy. This means
that every coherent state
$|\beta\rangle_{a_1}\langle\beta|\otimes\cdots\otimes|\beta\rangle_{a_{k}}\langle\beta|$
must saturate the bound~(\ref{croma}). To show that this is so, we
note that for any integer ${k}$ the matrices $G$ and $A$ have a null
eigenvalue (say for $j=1$), associated with the common eigenvector
$(1,1,\cdots,1)$.  In this case $e_1=0$ and $d_1=1/n$, so that
$\Theta_1=\openone_{j=1}$.  This means that for arbitrary
$|\varphi\rangle_{b_1}$, any state of the form
$R_\varphi\equiv|\varphi\rangle_{b_1}\langle\varphi|\otimes|0\rangle_{b_2}\langle
0|\otimes\cdots\otimes|0\rangle_{b_{k}}\langle 0|$ saturates the
bound~(\ref{croma}). If $|\varphi\rangle$ is not a coherent state, then
it corresponds to an entangled state of the $a_j$ modes, so it cannot be
written in the form
$\rho\otimes\cdots\otimes\rho$. Thus Tr$[R_\varphi \Theta]$
cannot be an output $k$-purity of the classical-noise channel. If, instead, we
repeat the same analysis of Eq.~(\ref{vuoto}) with
$|\varphi\rangle=|\sqrt{k}\beta\rangle$ being a coherent state, we 
find that the
resulting $R_\varphi$ is a tensor product of coherent states $|\beta\rangle$ in
the
$a_j$ modes, so that Tr$[R_\varphi \Theta]$ is the classical-noise
channel's output $k$-purity relative to the coherent-state input
$|\beta\rangle$.

\section{Derivation of lower bound {\bf 3)}}\labell{s:lb3}
In this appendix we derive the lower bound {\bf 3)}, given by
(\ref{lb3}) and~(\ref{lb3p}).

The $z$-purity Tr$[\rho^z]$ for $z\neq 1$ belongs to the class of
entropic measures defined in \cite{renyi}. Hence, for
$1<z'\leqslant z$, the state that minimizes Tr$[\rho^z]$ over the 
family of states
having constant Tr$[\rho^{z'}]=c$ is
known~\cite{renyi} to have a $q$-times degenerate eigenvalue
$\lambda_1$, and a nondegenerate eigenvalue
$\lambda_0=1-q\lambda_1\leqslant\lambda_1$. The value of the
parameters $\lambda_1$ and $q$ are determined by the constraint
\begin{eqnarray}
\lambda_0^{z'}+q\lambda_1^{z'}=c
\;,
\end{eqnarray}
which, for $1\geqslant\lambda_1\geqslant\lambda_0\geqslant 0$, gives
$q=\lfloor 1/\lambda_1\rfloor$, and can be written as
\begin{eqnarray}
h_{z'}(\lambda_1)
\equiv\left(1-\left\lfloor\frac
     1{\lambda_1}\right\rfloor\lambda_1\right)^{z'}
+\left\lfloor\frac
     1{\lambda_1}\right\rfloor\lambda_1^{z'}
=c
\;\labell{hdef},
\end{eqnarray}
where $\lfloor x\rfloor$ is the integer part of $x$. The function
$h_z(x)$ can be shown to be continuous and monotonically increasing
(see Fig.~\ref{f:hdef}), so that Eq.~(\ref{hdef}) has only one
solution in the range $c\in[0,1]$.  Hence, following \cite{renyi},
we can establish the inequality,
\begin{eqnarray}
\mbox{Tr}[\rho^z]\geqslant
h_{z}\left[h^{-1}_{z'}\left(\mbox{Tr}[\rho^{z'}]\right)\right]
\;,\labell{vittorio}
\end{eqnarray}
which applies for all $\rho$ and $z\geqslant z'>1$ ($h^{-1}$ being the
inverse of the function $h$). Because $h_{z}(h^{-1}_{z'}(x))$ is
monotonically increasing, Eq.~(\ref{vittorio}) can be recast
as
\begin{eqnarray}
S_{z'}(\rho)\geqslant-\frac{\ln\left[h_{z'}\left(h_z^{-1}
(\mbox{Tr}[\rho^z])\right)\right]}{z'-1}
\;.\labell{questaqui}
\end{eqnarray}
Evaluating this expression on the output states $\cn_n(\rho)$, we can
obtain a lower bound for $\ms_{z'}(\cn_n)$ by minimizing
both terms.  Moreover, we can replace the term Tr$[\rho^z]$ in
Eq.~(\ref{questaqui}) with its maximum value, because it is the argument
of a decreasing function. For $z=k$ an integer, we can then use the
results of Sec.~\ref{s:conjr} (where the maximum value of
Tr$\{[\cn_n(\rho)]^k\}$ was calculated) to derive (\ref{lb3})
from~(\ref{questaqui}).

The same analysis can be repeated for $z'<1$; in this case
$h_{z'}(x)$ is monotonically decreasing, which is compensated by
the sign change of the factor $1/(z'-1)$ in
Eq.~(\ref{questaqui}).

In order to derive (\ref{lb3p}), we apply
the analysis of \cite{renyi} to the von Neumann entropy $S(\rho)$
and Tr$[\rho^{z'}]$ with $z'<1$. Maximizing $S(\rho)$
over the family of states that have constant Tr$[\rho^{z'}]=c$, we
find that the optimal state has the same eigenvalue structure
$\{\lambda_0, \lambda_1\}$ encountered above. Equation~(\ref{vittorio})
is thus replaced by
\begin{eqnarray}
S(\rho)\leqslant
v\left[h^{-1}_{z'}\left(\mbox{Tr}[\rho^{z'}]\right)\right]
\;,\labell{vittorio1}
\end{eqnarray}
where
\begin{eqnarray}
v(x)\equiv-\left(1-\left\lfloor\frac 1x\right\rfloor x\right)
\ln\left(1-\left\lfloor\frac 1x\right\rfloor x\right)
-\left\lfloor\frac 1x\right\rfloor x\ln x
\nonumber\\\labell{vdef}
\end{eqnarray}
is the decreasing function plotted in Fig.~\ref{f:hdef}.  Because
$v\left[h^{-1}_{z'}\left(x\right)\right]$ is monotonically increasing,
Eq.~(\ref{vittorio1}) can be used to derive (\ref{lb3p}).

%da renyi.lb.f
\begin{figure}[hbt]
\begin{center}
\epsfxsize=1
\hsize\leavevmode\epsffile{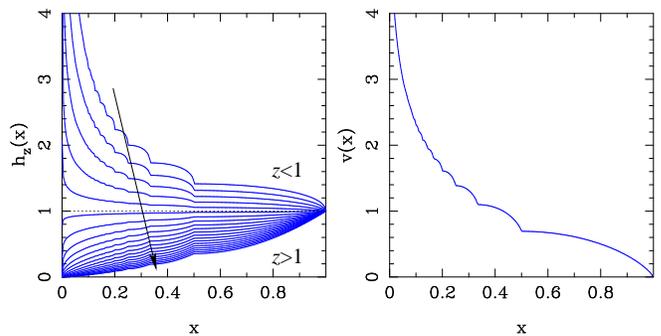}
\end{center}
%\vspace{-.5cm}
\caption{Left: plot of the function $h_z(x)$ from Eq.~(\ref{hdef}) as a
   function of $x$ for different values of $z$; $z$ increases from
   $1/2$ to $5/2$ in progressing along the direction of the arrow.
   For $z>1$, $h_z(x)$ is an increasing function, and for $z<1$ it is 
decreasing.
   Note that $h_z(1)=1$, $\lim_{x\to 0}h_z(x) = 0$ for $z>1$,
   and $\lim_{x\to 0}h_z(x) =\infty$ for $z<1$. Right: plot of the 
function $v(x)$
from
   Eq.~(\ref{vdef}).}  \labell{f:hdef}\end{figure}
%da renyi.lb.f

\vskip 1\baselineskip{\bf {Acknowledgments:}}
%\begin{acknowledgments}
   the Authors thank P. W. Shor, H. P. Yuen and P. Zanardi for useful
   discussions.  This work was funded by the ARDA, NRO, NSF, and by ARO
   under a MURI program.
%\end{acknowledgments}


\begin{references}
\bibitem{newnota} In the case ${z}=1$, Eq.~(\ref{fine1}) is
   trivially verified. However, this does not imply the
   conjecture in~(\ref{conjr}) owing to the presence of the $1/({z}-1)$
   factor in the R\'enyi entropy.
\end{references}
 \end{document}